\documentclass{article}

\usepackage{arxiv}
\usepackage{graphicx}
\usepackage{booktabs}
\usepackage{placeins}
\usepackage{amsmath}
\usepackage{microtype}
\usepackage{xcolor}
\usepackage{array}
\usepackage[colorlinks=true,allcolors=blue!55!black]{hyperref}
\usepackage[font=small,labelfont=bf]{caption}

\newcommand{\pt}{p_{\mathrm T}}

\newcolumntype{L}[1]{>{\raggedright\arraybackslash}p{#1}}

\title{Monte Carlo study of extending the BM@N tracking acceptance
toward midrapidity with the Vertex Silicon Plane}

\author{
D.~Dementev$^{1}$, D.~Baranov$^{3}$, V.~Leontyev$^{1,2}$,
Yu.~Murin$^{1}$, I.~Rufanov$^{1}$ \\[2mm]
\small $^{1}$Veksler and Baldin Laboratory of High Energy Physics,\\
\small Joint Institute for Nuclear Research, Dubna, Moscow Region, Russia\\
\small $^{2}$Moscow State University, Moscow, Russia\\
\small $^{3}$Meshcheryakov Laboratory of Information Technologies,\\
\small Joint Institute for Nuclear Research, Dubna, Moscow Region, Russia
}
\date{}

\hypersetup{
  pdftitle={Extending the BM@N tracking acceptance toward midrapidity with the Vertex Silicon Plane},
  pdfauthor={D. Dementev, V. Leontyev, Yu. A. Murin, D. Baranov},
  pdfsubject={VSP-assisted track reconstruction in the BM@N experiment},
  pdfkeywords={BM@N, silicon tracking, VSP, high-pT protons, tracking acceptance}
}
\renewcommand{\undertitle}{Research Note}
\renewcommand{\shorttitle}{BM@N VSP Research Note}
\renewcommand{\headeright}{}

\begin{document}
\maketitle

\begin{abstract}
\noindent
The Vertex Silicon Plane (VSP), installed close to the target in BM@N
Run~9, can extend the tracking acceptance of the central tracking system toward larger polar angles.
Using Monte Carlo simulations of Xe+Sn collisions at
$3\,A\mathrm{GeV}$, we reconstruct short tracklets from the primary
vertex, one VSP hit, and hits in the first two Gas Electron Multiplier
(GEM) stations. These tracklets extend the reconstructed primary-proton
coverage toward midrapidity and higher transverse
momenta.

For the \(10\)--\(30\%\) centrality interval of the independent validation
dataset, the conditional reconstruction efficiency for reconstructible
high-\(\pt\) primary protons after BDT selection is \(67.0\%\), while the
tracklet purity reaches \(86.5\%\). For
\(p_{\mathrm T}>1.2\,\mathrm{GeV}/c\), the VSP-assisted
tracklets increase the number of reconstructed primary protons by \(17.5\%\)
relative to standard BM@N tracking. Of these additional protons, \(84.6\%\)
lie within \(|y|<0.3\), where the relative reconstruction gain reaches
\(45.5\%\).

The added acceptance is concentrated in the high-$p_{\mathrm T}$,
near-midrapidity region relevant for identified-proton observables.
Its use in data requires particle identification in the same region.
\end{abstract}

\keywords{BM@N \and Vertex Silicon Plane \and track reconstruction}

\section{Introduction}

The BM@N (Baryonic Matter at Nuclotron) fixed target experiment at the NICA accelerator
complex studies nucleus--nucleus collisions with heavy-ion systems extending
to Bi+Bi and beam kinetic energies up to about
$4.5\,A\mathrm{GeV}$~\cite{BMNspectrometer2024,Senger2019}. In this energy
range, heavy-ion collisions provide access to strongly interacting matter at
high net-baryon density, and proton observables are particularly useful for
studying the collective response of the system and the equation of state
(EoS) of dense nuclear matter.

Transport calculations indicate that high-momentum nucleons emitted near
midrapidity are especially sensitive to the momentum dependence of the
nuclear mean field at supra-saturation densities
~\cite{Danielewicz2000}. This makes the high-$p_{\mathrm T}$ region near
midrapidity particularly useful for constraining the mean-field dynamics of
the high-density stage. Measurements of proton anisotropic flow at comparable collision
energies show strong dependences on both transverse momentum and rapidity and
provide important constraints for comparisons with transport calculations
~\cite{LeFevre2016,HADES2023,STAR2022}.

Midrapidity is particularly important for these observables because it
selects the participant region around the nucleon--nucleon center-of-mass
frame rather than the projectile- or target-fragmentation regions. The
slope of directed flow is conventionally evaluated around $y=0$, where
$y=y_{\rm lab}-y_{\rm cm}$, while measurements at comparable collision
energies show
that the magnitude of proton elliptic flow is largest around midrapidity and
varies strongly with $p_{\mathrm T}$~\cite{HADES2023}. Recent BM@N
measurements of the directed flow of identified protons in Xe+CsI
collisions at $3.8\,A\mathrm{GeV}$ provide a direct experimental benchmark
for such rapidity-dependent proton-flow observables at Nuclotron energies
~\cite{Mamaev2026}.

The experimental sensitivity to these observables depends not only on the
available event statistics but also on the phase-space coverage of the
tracking system. The combination of high $p_{\mathrm T}$ and
near-midrapidity is therefore a particularly relevant target: the former
enhances sensitivity to the early, dense stage of the reaction, while the
latter selects the participant region in which collective-flow observables
are conventionally characterized. The standard BM@N tracking configuration
is optimized for the forward spectrometer acceptance and has limited
coverage at larger polar angles. Consequently, extending the reconstructed
$(p_{\mathrm T},y)$ phase space can provide access to proton trajectories
that are poorly represented in the standard reconstructed sample.

During BM@N Run~9, the Vertex Silicon Plane (VSP) was installed close to
the interaction region, primarily to validate the free-streaming silicon
readout chain under realistic heavy-ion running conditions. Owing to its
near-target position and its geometrical overlap with the first Gas Electron
Multiplier (GEM) stations, the VSP provides an additional tracking
constraint for particles emitted at larger polar angles. This overlap allows
short tracklets to be reconstructed from the primary
vertex (PV) as their first trajectory point, one VSP hit, and measurements in
the first two GEM stations.
The resulting acceptance extension is expected to populate the region toward
midrapidity and higher transverse momenta.

Reconstructing such tracklets is challenging because the trajectory is
constrained by only a few spatial measurements, the VSP and the first GEM
station are separated by roughly $40~\mathrm{cm}$, and the high hit
multiplicity in central heavy-ion collisions produces a large combinatorial
background. This work evaluates to what
extent the additional geometrical acceptance provided by the VSP can be
retained after hit reconstruction, magnetic-field propagation, and
ambiguity resolution. A dedicated tracklet reconstruction algorithm is
developed and evaluated using Monte Carlo Xe+Sn events at
$3\,A\mathrm{GeV}$, and its contribution is compared directly with the
standard reconstructed proton sample in the $(p_{\mathrm T},y)$ phase space.

This study addresses tracking acceptance and reconstruction performance.
Monte Carlo truth information is used to identify the proton component in
this performance study. Applying the method to identified-proton observables
in data, including anisotropic flow, therefore requires particle
identification (PID) in the corresponding extended-acceptance region.

\section{Simulation sample and tracklet reconstruction}

\subsection{Monte Carlo dataset}

The simulated events were drawn from a production of $10^{7}$ minimum-bias
Xe+Sn collisions at a beam kinetic energy of $3\,A\mathrm{GeV}$ generated
with DCM--SMM~\cite{DCMSMM2020}. Detector transport, digitization, and
reconstruction were performed within the BmnRoot framework using the Run~9
detector geometry. The present analysis uses $75\,000$ events, divided into two
statistically independent training and validation datasets containing
approximately $50\,000$ and $25\,000$ events, respectively.

The quantitative reconstruction performance is evaluated on the independent
validation dataset restricted to the $10$--$30\%$ centrality interval,
corresponding to a simulated impact parameter $4.1 \le b < 7.1~\mathrm{fm}$.

\subsection{VSP--GEM tracklet finder}

The VSP consists of six double-sided silicon strip detector modules located
approximately $7.4$--$8.5~\mathrm{cm}$ downstream from the target and covering
polar angles of approximately $23^{\circ}$--$52^{\circ}$. Each module provides
a two-dimensional hit measurement. The detector technology and the streaming
readout implementation are described in Refs.~\cite{VSPbeam2025,VSPstation2026}.

\begin{figure}[t]
\centering
\begin{minipage}[t]{0.49\textwidth}
    \centering
    \includegraphics[width=\linewidth]{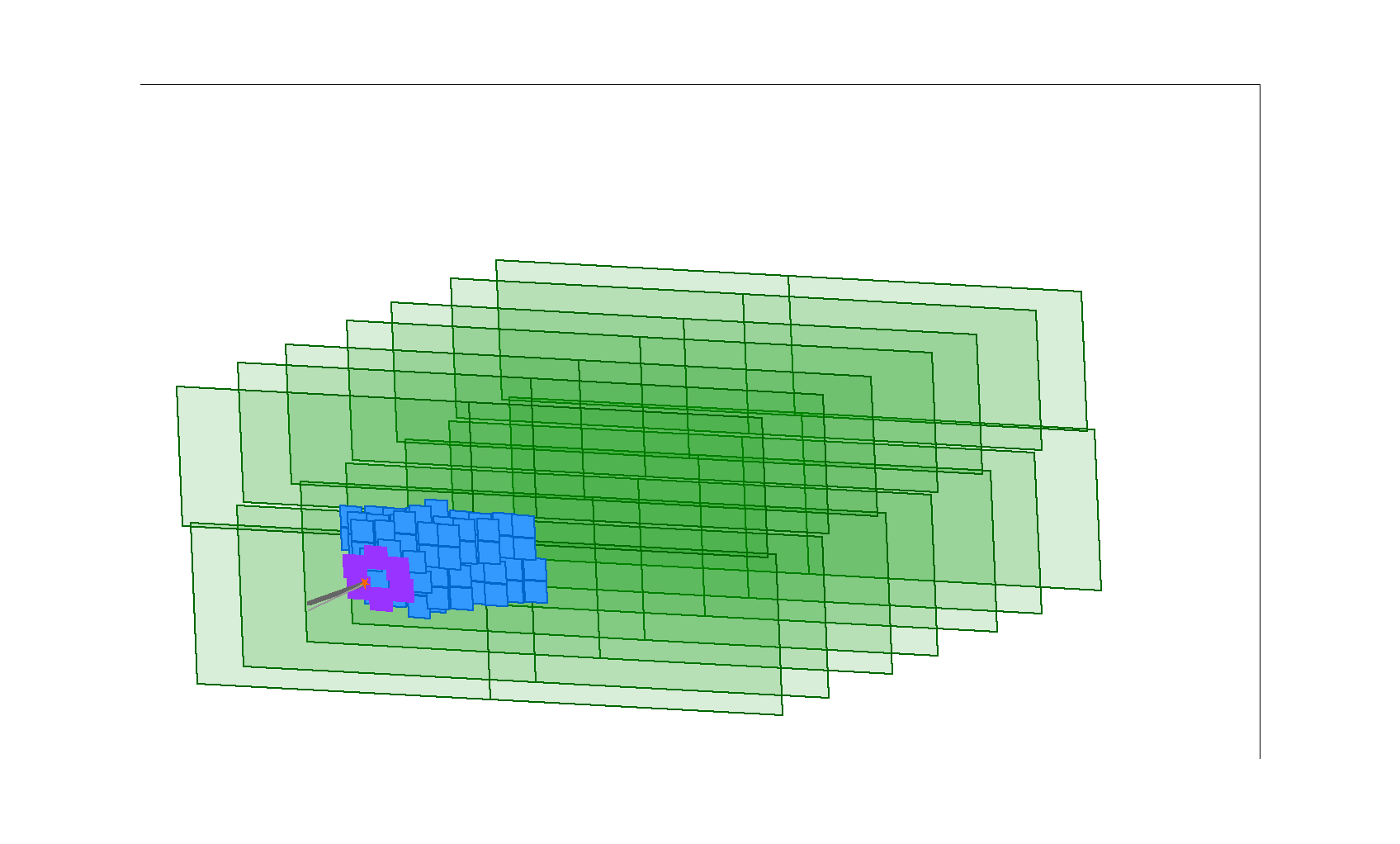}\par
    \vspace{1mm}
    \textbf{(a)}
\end{minipage}
\hfill
\begin{minipage}[t]{0.49\textwidth}
    \centering
    \includegraphics[width=\linewidth]{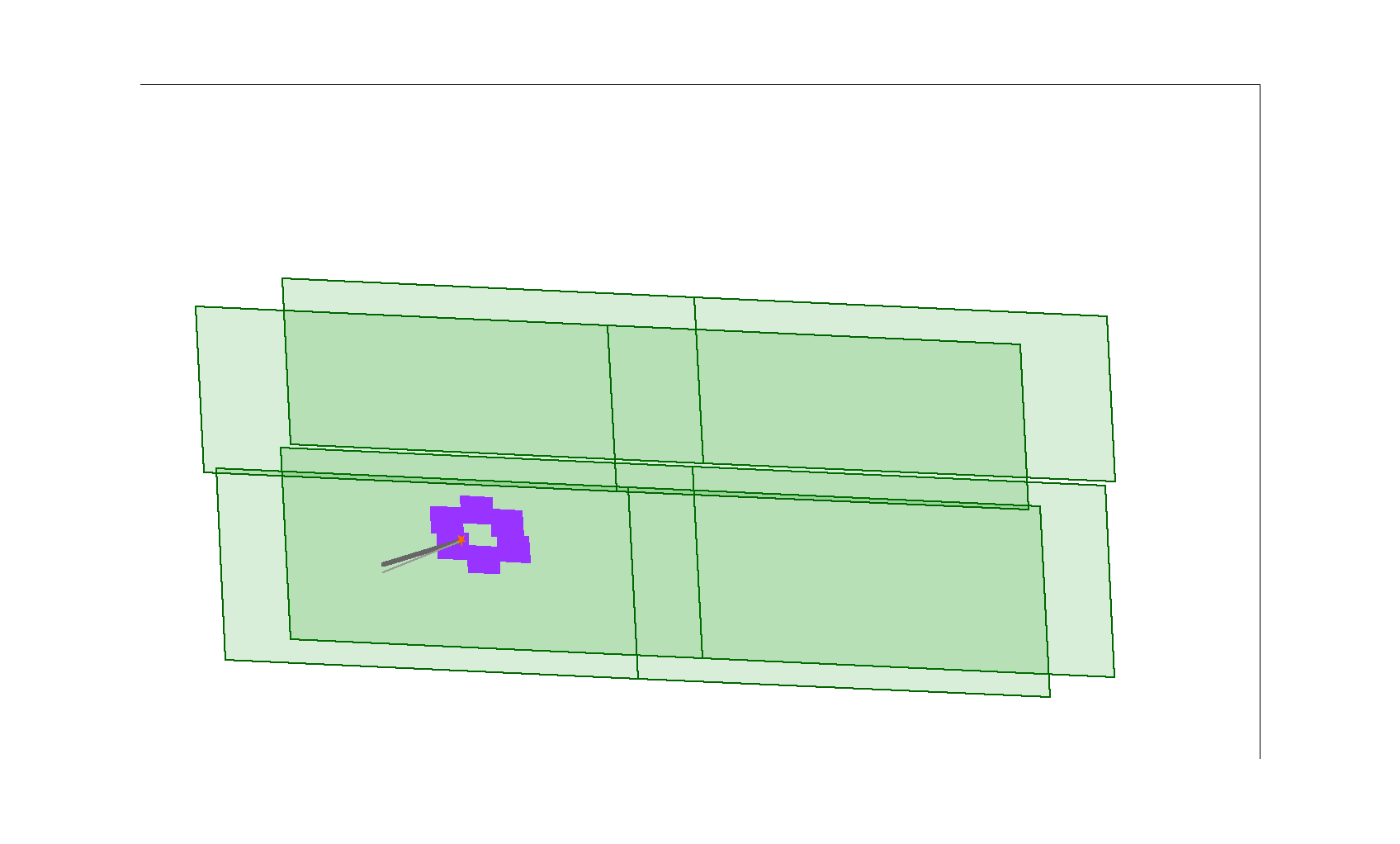}\par
    \vspace{1mm}
    \textbf{(b)}
\end{minipage}
\caption{Geometry-based views of the BM@N Run~9 tracking configuration.
(a) Full tracking system. (b) Detector subset used for the VSP--GEM
tracklet reconstruction: the reconstructed primary vertex (PV), used as the
first trajectory point, the near-target VSP station, and the first two
downstream GEM stations.}
\label{fig:geom}
\end{figure}

The VSP-assisted reconstruction uses short tracklets built from the
reconstructed primary vertex as the first trajectory point, one VSP hit, and
measurements in the first two GEM stations, GEM1 and GEM2. In the following,
these reconstructed objects are referred to simply as tracklets.
Figure~\ref{fig:geom} shows the full Run~9 tracking geometry and, for
comparison, the reduced detector subset used in the VSP--GEM tracklet
reconstruction.

A dedicated sequential finder was developed for the VSP-assisted
tracklets. The vertex--VSP seed is propagated through the magnetic field
of the SP-41 analyzing magnet. Candidate GEM1 hits compatible with the
propagated state are considered, and the track parameters are updated with
a Kalman filter (KF) before propagation to GEM2. The two physical
half-planes of each GEM station are treated at their individual
$z$ positions.

The initial inverse-rigidity scan covers positive values in the range
$0.05 \le q/p \le 2.5~(\mathrm{GeV}/c)^{-1}$ in steps of
$0.05~(\mathrm{GeV}/c)^{-1}$. Candidate combinations are ordered by a
geometrical quality measure. If two candidates share a detector hit, only the
higher-ranked candidate is retained. Magnetic-field propagation follows the
procedure used in the BM@N Vector Finder~\cite{Zinchenko2024}, together with
the standard Kalman-filter formalism~\cite{Fruehwirth1987}.

\subsection{BDT candidate selection}

To suppress fake tracklet candidates remaining after the geometrical
ambiguity selection, a gradient-boosted decision tree (BDT) implemented
in the ROOT Toolkit for Multivariate Data Analysis (TMVA) is applied as a
post-reconstruction quality filter~\cite{Hoecker2007,Friedman2001}.
The classifier uses 24 reconstruction-level observables describing the
sequential GEM1 and GEM2 search, the KF updates, the consistency of the fitted
track parameters, the residuals at the three detector stations, the track
direction, and the primary-vertex track multiplicity.

Monte Carlo truth is used only to define the training labels. A candidate is
labelled as signal when its VSP, GEM1, and GEM2 hits are associated with the
same MC particle; all other combinations are labelled as background. No
particle-species requirement is included in this definition.

The classifier was trained on the training dataset and evaluated on the
statistically independent validation dataset. The validation dataset was not
used during the training or optimization of the classifier. The area under
the validation ROC curve is 0.872. The working point $S_{\rm BDT}>-0.10$ was fixed as a
compromise between tracklet purity and high-$p_{\mathrm T}$ proton retention.
The numerical finder settings, complete input-variable list, and BDT
configuration are given in Appendix~\ref{app:configuration}.

\subsection{Performance definitions}

In this study, the high-$p_{\mathrm T}$ region is defined by
$p_{\mathrm T}>1.2~\mathrm{GeV}/c$. For free elastic
$NN\rightarrow NN$ scattering of a beam nucleon with kinetic energy
$T_{\rm beam}=3~\mathrm{GeV}$ on a stationary nucleon, the nucleon momentum
in the center-of-mass frame is
$p^{*}=\sqrt{m_{N}T_{\rm beam}/2}\simeq1.19~\mathrm{GeV}/c$. Since
transverse momentum is invariant under a longitudinal boost and satisfies
$p_{\mathrm T}\leq p^{*}$ for this elementary two-body kinematics, the
chosen threshold corresponds approximately to the upper transverse-momentum
scale of free elastic $NN$ scattering. Nuclear Fermi motion, rescattering,
and collective mean-field dynamics smear this scale in nucleus--nucleus
collisions~\cite{Danielewicz2000}.

The tracklet-reconstruction efficiency is evaluated for primary protons
satisfying the high-$p_{\mathrm T}$ condition defined from the MC momentum,
with a valid reconstructed primary vertex and truth-linked reconstructed hits
in the VSP, GEM1, and GEM2. A proton contributes to the numerator when the
corresponding truth-matched tracklet is selected.
The quoted efficiency is therefore conditional on the presence of
reconstructed hits in all three tracking stations and characterizes the
pattern-recognition efficiency of the tracklet finder.

\FloatBarrier

\section{Results}

\subsection{Extension of the reconstructed phase space}

\begin{figure}[t]
\centering
\includegraphics[width=0.98\textwidth]{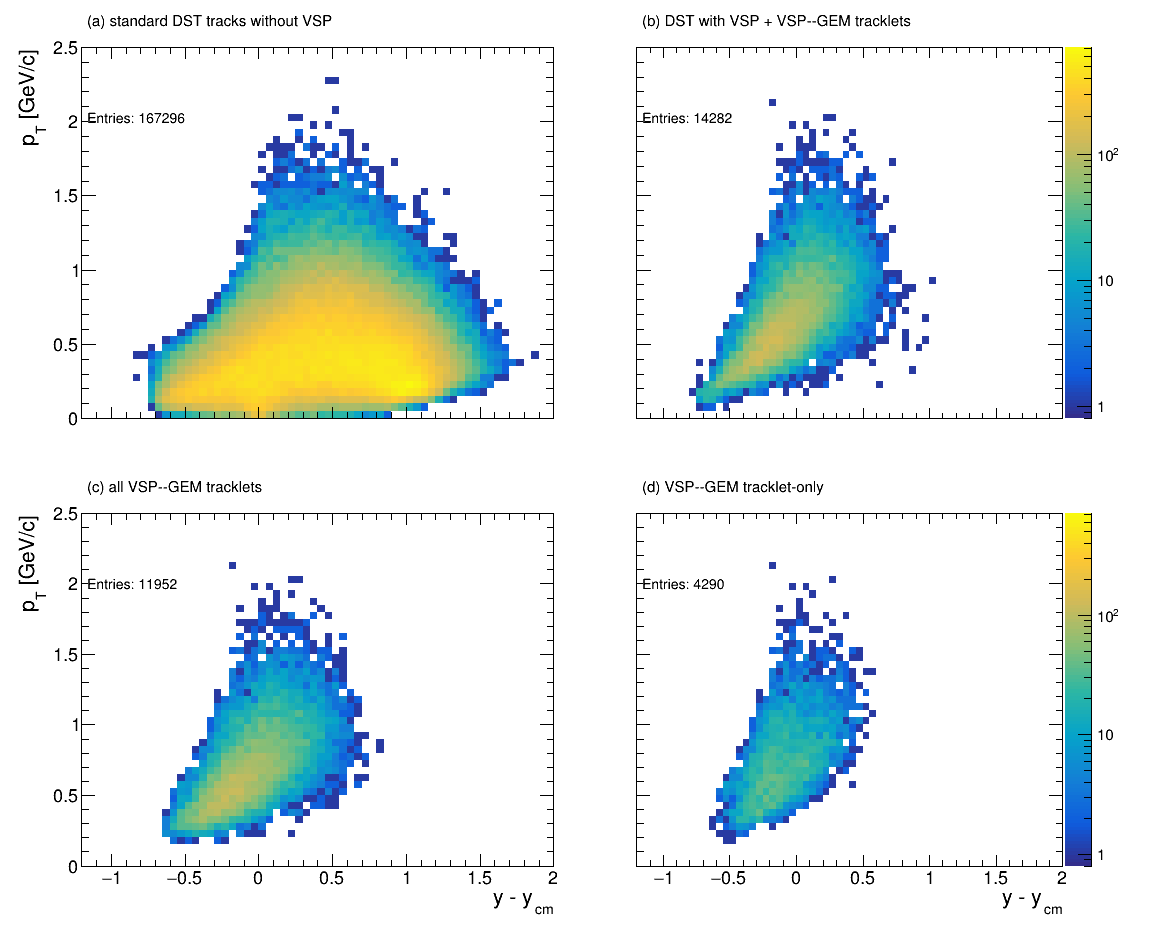}
\caption{Reconstructed primary-proton phase space for the $10$--$30\%$
centrality interval of the validation dataset in Xe+Sn collisions at
$3\,A\mathrm{GeV}$, shown in the $(y,p_{\mathrm T})$ phase
space. (a) Primary protons reconstructed by the standard BM@N offline tracking
without a linked VSP hit. (b) Combined VSP-sensitive sample consisting of
standard tracks with a linked VSP hit and BDT-selected VSP--GEM
tracklets. (c) All BDT-selected VSP--GEM tracklets. (d) VSP--GEM tracklets
corresponding to primary protons not reconstructed by the standard offline
tracking. MC-truth kinematics are used for the truth-linked reconstructed
objects, so that the comparison reflects reconstruction coverage
independently of momentum-resolution effects. The number of unique
reconstructed primary protons is indicated in each panel.}
\label{fig:recogain}
\end{figure}

Figure~\ref{fig:recogain} compares the primary-proton phase-space populations
reconstructed by the standard BM@N offline tracking and by the VSP--GEM
tracklet finder. It distinguishes standard tracks without a linked VSP hit,
standard tracks with a linked VSP hit, and dedicated VSP--GEM tracklets.

For the high-$p_{\mathrm T}$ region defined above, the tracklet-only component
adds 318 primary protons to the 1822 reconstructed by the standard BM@N
tracking, corresponding to a $17.5\%$ increase. Of these 318 additional
protons, 269 ($84.6\%$) lie within $|y|<0.3$. In this interval, the
standard tracking reconstructs 591 protons; adding the VSP--GEM tracklets
raises the reconstructed population to 860, corresponding to a $45.5\%$
increase. The acceptance gain is therefore strongly concentrated near
midrapidity.

\FloatBarrier

\subsection{Tracklet reconstruction efficiency}

For the $10$--$30\%$ centrality interval of the validation dataset, the
dedicated sequential tracklet finder reconstructs $76.3\%$ of the
reconstructible high-$p_{\mathrm T}$ primary protons. This value characterizes the
conditional reconstruction performance before the multivariate quality selection.

After applying the BDT selection, the conditional reconstruction efficiency is
$67.0\%$. The corresponding retention of correctly reconstructed
high-$p_{\mathrm T}$ proton tracklets is $87.9\%$.

\FloatBarrier

\subsection{Tracklet purity and BDT selection}

The quality of the reconstructed tracklet population is characterized by
the tracklet purity,
\[
  P_{\rm trk} =
  \frac{N_{\rm truth\text{-}matched}}{N_{\rm selected}},
\]
where $N_{\rm selected}$ is the total number of tracklets at the selection
stage under consideration, and $N_{\rm truth\text{-}matched}$ is the subset
whose VSP, GEM1, and GEM2 hits are associated with the same MC particle. The
fake fraction is $1-P_{\rm trk}$.

For the $10$--$30\%$ centrality interval of the validation dataset, before
the BDT selection the tracklet purity is $71.4\%$, while $28.6\%$ of the
selected candidates are fake. This indicates that
combinatorial ambiguities remain significant after the geometrical
selection alone.

Applying the BDT working point $S_{\rm BDT}>-0.10$ increases the tracklet
purity to $86.5\%$ and reduces the fake fraction to $13.5\%$. At the same
time, $89.3\%$ of the truth-matched tracklets are retained. The BDT rejects
$65.3\%$ of the fake candidates present before the multivariate selection.

\begin{table}[t]
\centering
\caption{Tracklet-selection performance for the $10$--$30\%$ centrality
interval of the validation dataset. The BDT working point is
$S_{\rm BDT}>-0.10$.}
\label{tab:ml}
\small
\begin{tabular}{lcc}
\toprule
Quantity & before BDT & after BDT \\
\midrule
Selected tracklets                & 23\,167 & 17\,070 \\
Truth-matched tracklets           & 16\,534 & 14\,771 \\
Tracklet purity                   & 71.4\%  & 86.5\% \\
Fake fraction                     & 28.6\%  & 13.5\% \\
Truth-matched-tracklet retention  & --      & 89.3\% \\
High-$p_{\mathrm T}$ proton retention
                                  & --      & 87.9\% \\
\bottomrule
\end{tabular}
\end{table}

\begin{figure}[t]
\centering
\includegraphics[width=0.98\textwidth]{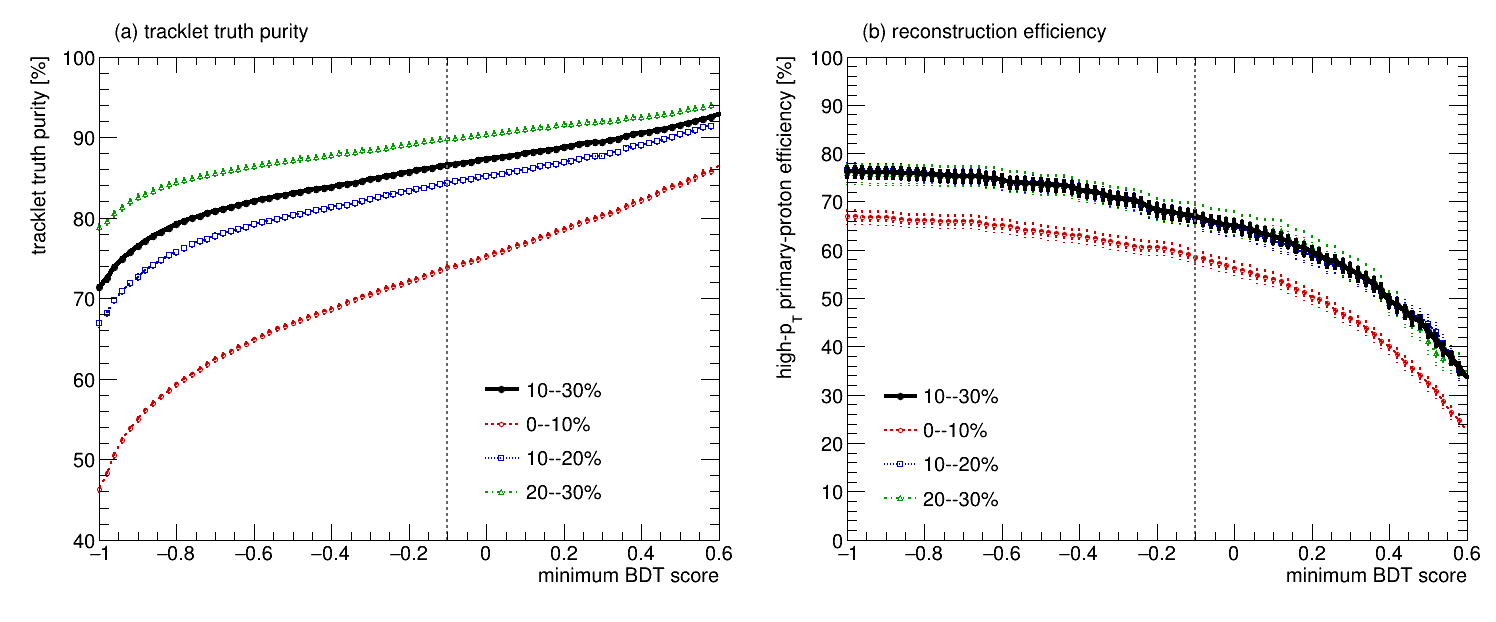}
\caption{Dependence of (a) the tracklet truth purity and (b) the
high-$p_{\mathrm T}$ primary-proton reconstruction efficiency on the
minimum BDT score for several centrality intervals of the held-out
validation dataset. The $10$--$30\%$ interval used for the main analysis
is highlighted by the thick black curve. The vertical dashed line marks
the working point $S_{\rm BDT}>-0.10$.}
\label{fig:bdtcent}
\end{figure}

The balance between tracklet purity and reconstruction efficiency depends
on event centrality, as shown in Fig.~\ref{fig:bdtcent}. The largest
combinatorial penalty is observed for the most central $0$--$10\%$ class,
where the tracklet purity is substantially lower over the full BDT-score
range. The efficiency of reconstructible high-$p_{\mathrm T}$ primary
protons varies more moderately with centrality. The $10$--$30\%$ interval
used in the subsequent analysis therefore provides a practical compromise
between event statistics, tracklet purity and signal efficiency.

\FloatBarrier

The BDT selection therefore substantially improves the pattern-recognition
quality while retaining most of the truth-matched tracklets. The particle
composition of the selected tracklet population is illustrated by the
fitted inverse-rigidity distribution in Fig.~\ref{fig:qpcomp}.

The selected population is dominated by protons in MC truth, with smaller
contributions from deuterons and positive pions. The $q/p$ distribution is
used here only as a composition diagnostic: the magnetic fit determines
the charge-to-momentum ratio but does not provide particle identification.

\subsection{Transverse-momentum reconstruction}

\begin{figure}[t]
\centering
\begin{minipage}[t]{0.48\textwidth}
  \centering
  \includegraphics[width=\linewidth]{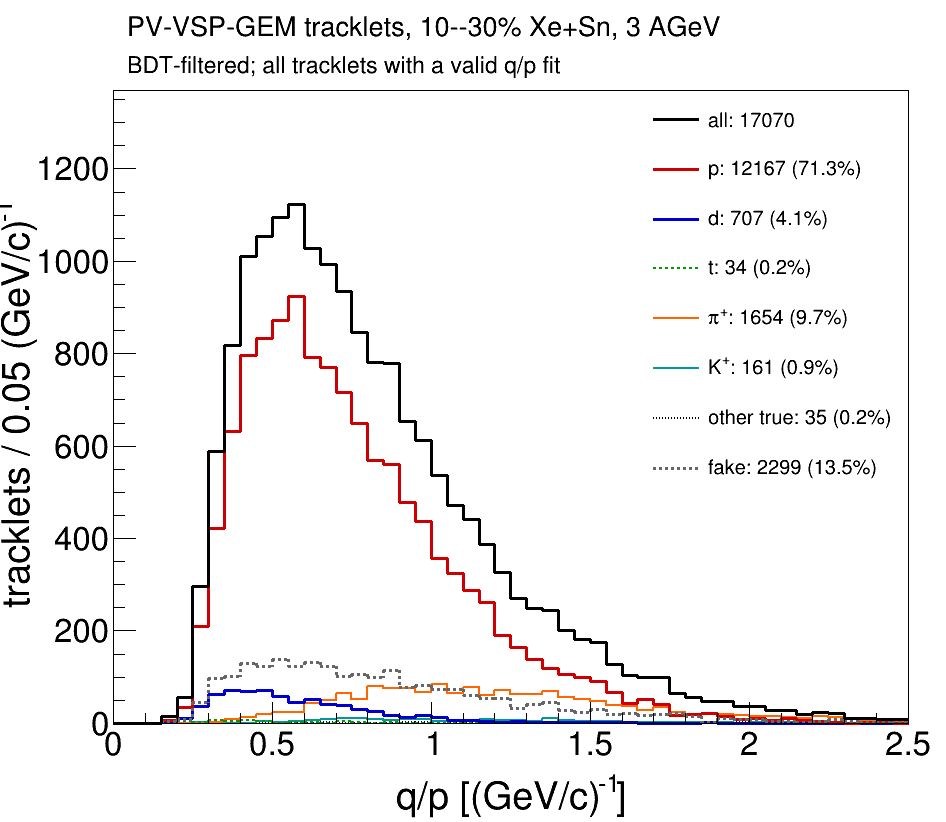}
  \caption{Fitted inverse-rigidity $q/p$ distribution for BDT-selected
  tracklets in the $10$--$30\%$ centrality interval of the validation
  dataset in Xe+Sn collisions at $3\,A\mathrm{GeV}$. The black contour
  shows all selected tracklets with a valid magnetic fit. Colored contours
  indicate the principal MC-truth particle species, while the grey dashed
  curve shows fake tracklets. Protons account for $71.3\%$ of the inclusive
  fitted population, and the fake fraction is $13.5\%$.}
  \label{fig:qpcomp}
\end{minipage}
\hfill
\begin{minipage}[t]{0.48\textwidth}
  \centering
  \includegraphics[width=\linewidth]{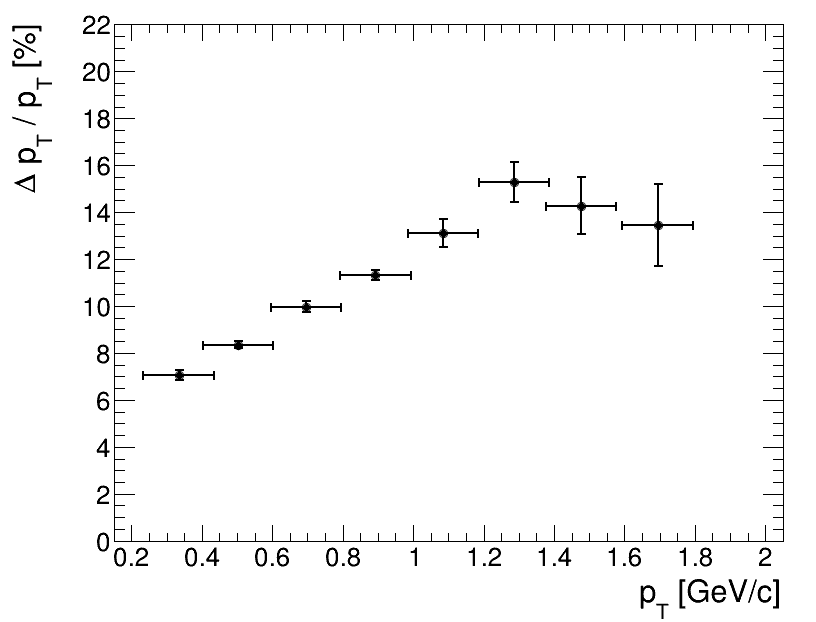}
  \caption{Relative transverse-momentum resolution for truth-matched
  primary-proton tracklets in the $10$--$30\%$ centrality interval of the
  validation dataset after the BDT quality selection. The points show the
  RMS68 width of $\Delta p_{\mathrm T}/p_{\mathrm T}^{\rm true}$ in bins of
  $p_{\mathrm T}^{\rm true}$. Horizontal bars indicate the bin width and
  vertical bars the statistical uncertainty of the RMS68 estimator.}
  \label{fig:ptres}
\end{minipage}
\end{figure}

The transverse-momentum resolution of the reconstructed tracklets is shown
in Fig.~\ref{fig:ptres}. The resolution is characterized by the RMS68 width
of the relative residual
\[
  \frac{\Delta p_{\mathrm T}}{p_{\mathrm T}^{\rm true}}
  =
  \frac{p_{\mathrm T}^{\rm reco}-p_{\mathrm T}^{\rm true}}
       {p_{\mathrm T}^{\rm true}},
\]
evaluated in bins of true transverse momentum. The RMS68 is defined as
half the width of the central $68\%$ interval of the residual distribution
and is used as a robust measure of the resolution with reduced sensitivity
to non-Gaussian tails.

The relative resolution is approximately $7\%$ at low transverse momentum
and increases to about $13$--$15\%$ in the high-$p_{\mathrm T}$ region.
No selection on reconstructed $p_{\mathrm T}$ is applied in this
evaluation, so migration across the high-$p_{\mathrm T}$ boundary is
treated as a momentum-resolution effect rather than as a tracking
inefficiency.

\FloatBarrier

\section{Conclusion}

The VSP provides a reconstructable extension of the BM@N tracking coverage
in the $(p_{\mathrm T},y)$ phase space, concentrated specifically toward
midrapidity and higher transverse momenta. The high-$p_{\mathrm T}$ threshold
$p_{\mathrm T}>1.2~\mathrm{GeV}/c$ corresponds approximately to the
upper transverse-momentum scale of free elastic $NN$ scattering at the
studied beam energy. For the $10$--$30\%$ centrality interval of the independent validation
dataset, the BDT-selected tracklet-only component increases the reconstructed
primary-proton population by $17.5\%$ in this region.

The acceptance gain is strongly localized in rapidity: 269 of the 318
additional high-$p_{\mathrm T}$ primary protons are found within
$|y|<0.3$. In this interval, the VSP--GEM tracklets increase the standard
sample from 591 to 860 protons, corresponding to a $45.5\%$ gain, compared
with $17.5\%$ when integrated over the full rapidity range.

The dedicated sequential tracklet finder reconstructs $76.3\%$ of
reconstructible high-$p_{\mathrm T}$ primary protons. After the BDT quality
selection, the conditional efficiency is $67.0\%$, with $87.9\%$ retention of correctly
reconstructed high-$p_{\mathrm T}$ proton tracklets. At the same time, the
tracklet purity increases from $71.4\%$ to $86.5\%$. The relative
transverse-momentum resolution is approximately $7$--$15\%$ over the
studied momentum range.

The demonstrated gain extends the BM@N tracking acceptance into the
intersection of two physically important regions: high transverse momentum,
where transport calculations predict enhanced sensitivity to the dense-stage
mean field, and midrapidity, where participant-region collective flow is
most directly characterized. The present study establishes
the tracking capability there; its use for identified-proton measurements
requires particle identification in the corresponding extended-acceptance region.

\appendix
\section{Reconstruction and BDT configuration}
\label{app:configuration}

The parameters below correspond to the fixed configuration used for all
results presented in this work; no claim of a globally optimal
hyperparameter choice is made. The positive $q/p$ interval reflects the focus
on positively charged particles. The geometrical requirements define the
candidate preselection, while the remaining combinatorial background is
suppressed by the BDT selection.

\begin{table}[htbp]
\centering
\caption{Configuration of the sequential VSP--GEM tracklet finder.}
\label{tab:finder_configuration}
\small
\begin{tabular}{L{0.48\textwidth}L{0.40\textwidth}}
\toprule
Parameter & Value \\
\midrule
Positive-$q/p$ scan range
  & $0.05$--$2.5~(\mathrm{GeV}/c)^{-1}$ \\
$q/p$ scan step
  & $0.05~(\mathrm{GeV}/c)^{-1}$ \\
GEM1 search-corridor parameters
  & $1.2~\mathrm{cm}$ in $x$; $0.8~\mathrm{cm}$ in $y$ \\
GEM2 search-corridor parameters
  & $1.6~\mathrm{cm}$ in $x$; $1.0~\mathrm{cm}$ in $y$ \\
Retained hypotheses per GEM1 candidate
  & up to 20 \\
PV--VSP shooting iterations
  & up to 4 \\
Final PV--VSP shooting residual
  & below $0.002~\mathrm{cm}$ \\
Separation of the GEM half-planes in $z$
  & $0.2~\mathrm{cm}$ \\
Geometrical candidate preselection
  & $\chi^{2}_{\rm rank}<10$ \\
Hard cut on individual GEM KF $\chi^{2}$ contributions
  & none \\
Ambiguity resolution
  & geometrical-rank ordering followed by no-shared-hit selection \\
\bottomrule
\end{tabular}
\end{table}

\begin{table}[htbp]
\centering
\caption{Reconstruction-level variables used as BDT inputs. Branch names are
given as used in the training tree.}
\label{tab:bdt_inputs}
\small
\begin{tabular}{L{0.27\textwidth}L{0.61\textwidth}}
\toprule
Variable group & Input branches \\
\midrule
Geometrical rank
  & \texttt{rankChi2} \\
Sequential search
  & \texttt{seqScoreG1}, \texttt{seqScoreG2}, \texttt{seqDx1},
    \texttt{seqDy1}, \texttt{seqDx2}, \texttt{seqDy2} \\
KF updates and rigidity
  & \texttt{kfChi2G1}, \texttt{kfChi2G2}, \texttt{seedQp},
    \texttt{kfQpAfterG1} \\
Track slopes and consistency
  & \texttt{kfTxAfterG1}, \texttt{kfTyAfterG1}, \texttt{dtxPair},
    \texttt{dtyPair}, \texttt{dyMid} \\
Final-fit residuals
  & \texttt{rxV}, \texttt{ryV}, \texttt{rx1}, \texttt{ry1},
    \texttt{rx2}, \texttt{ry2} \\
Track direction and PV multiplicity
  & \texttt{thetaDeg}, \texttt{pvNTracks} \\
\bottomrule
\end{tabular}
\end{table}

\begin{table}[htbp]
\centering
\caption{Configuration and sample composition of the TMVA classifier.}
\label{tab:bdt_configuration}
\small
\begin{tabular}{L{0.48\textwidth}L{0.40\textwidth}}
\toprule
Parameter & Value \\
\midrule
TMVA method & BDTG \\
Number of input variables & 24 \\
Spectator variables & none \\
Number of trees & 600 \\
Boost type & gradient boosting \\
Shrinkage & 0.05 \\
Bagged sample fraction & 0.60 \\
Number of candidate split points & 30 \\
Maximum tree depth & 3 \\
Minimum node size & 2.5\% \\
Explicit variable transformations & none \\
Training normalization & equal total signal and background weights \\
Training sample & approximately $50\,000$ events \\
Training candidates & 58\,676 signal; 44\,645 background \\
Validation sample & approximately $25\,000$ events \\
Validation candidates & 29\,073 signal; 21\,196 background \\
Validation ROC area & 0.872 \\
Working point & $S_{\rm BDT}>-0.10$ \\
\bottomrule
\end{tabular}
\end{table}

\FloatBarrier


\begin{thebibliography}{99}\small

\bibitem{BMNspectrometer2024}
S.~Afanasiev \emph{et al.},
\emph{The BM@N spectrometer at the NICA accelerator complex},
Nucl. Instrum. Meth. A \textbf{1065} (2024) 169532,
doi:10.1016/j.nima.2024.169532.

\bibitem{Senger2019}
P.~Senger \emph{et al.},
\emph{Upgrading the Baryonic Matter at the Nuclotron Experiment at NICA
for Studies of Dense Nuclear Matter},
Particles \textbf{2} (2019) 481--490,
doi:10.3390/particles2040029.


\bibitem{Danielewicz2000}
P.~Danielewicz,
\emph{Determination of the Mean-Field Momentum-Dependence using Elliptic Flow},
Nucl. Phys. A \textbf{673} (2000) 375--410,
doi:10.1016/S0375-9474(00)00083-X.

\bibitem{LeFevre2016}
A.~Le F\`evre, Y.~Leifels, W.~Reisdorf, J.~Aichelin and Ch.~Hartnack,
\emph{Constraining the nuclear matter equation of state around twice saturation density},
Nucl. Phys. A \textbf{945} (2016) 112--133,
doi:10.1016/j.nuclphysa.2015.09.015.

\bibitem{HADES2023}
HADES Collaboration,
\emph{Proton, deuteron and triton flow measurements in Au+Au collisions at $\sqrt{s_{NN}}=2.4$ GeV},
Eur. Phys. J. A \textbf{59} (2023) 80,
doi:10.1140/epja/s10050-023-00936-6.

\bibitem{STAR2022}
STAR Collaboration,
\emph{Disappearance of partonic collectivity in $\sqrt{s_{NN}}=3$ GeV Au+Au collisions at RHIC},
Phys. Lett. B \textbf{827} (2022) 137003,
doi:10.1016/j.physletb.2022.137003.

\bibitem{Mamaev2026}
M.~Mamaev and A.~Taranenko,
\emph{Measuring the directed flow of protons in Xe+CsI collisions at the energy of $3.8A$ GeV at the BM@N experiment},
Int. J. Mod. Phys. E (2026),
doi:10.1142/S0218301326410132.

\bibitem{DCMSMM2020}
M.~Baznat, A.~Botvina, G.~Musulmanbekov, V.~Toneev and V.~Zhezher,
\emph{Monte-Carlo generator of heavy ion collisions DCM-SMM},
Phys. Part. Nucl. Lett. \textbf{17} (2020) 303--324,
doi:10.1134/S1547477120030024,
arXiv:1912.09277 [nucl-th].

\bibitem{VSPbeam2025}
D.~Dementev \emph{et al.},
\emph{In-beam tests of double-sided silicon strip detector modules for the
BM@N experiment},
Nucl. Instrum. Meth. A \textbf{1075} (2025) 170390,
doi:10.1016/j.nima.2025.170390.

\bibitem{VSPstation2026}
D.~Dementev \emph{et al.},
\emph{Silicon Tracking Station with a Streaming Readout as a Part of the
BM@N Tracking System},
Phys. Part. Nucl. \textbf{57} (2026) 379--383,
doi:10.1134/S1063779625701424.

\bibitem{Zinchenko2024}
D.~A.~Zinchenko \emph{et al.},
\emph{Development of the Vector Finder Toolkit for Track Reconstruction in the BM@N Experiment},
Phys. Part. Nucl. Lett. \textbf{21} (2024) 544--552,
doi:10.1134/S1547477124700560.

\bibitem{Fruehwirth1987}
R.~Fr\"uhwirth,
\emph{Application of Kalman filtering to track and vertex fitting},
Nucl. Instrum. Meth. A \textbf{262} (1987) 444--450,
doi:10.1016/0168-9002(87)90887-4.


\bibitem{Hoecker2007}
A.~Hoecker, P.~Speckmayer, J.~Stelzer, J.~Therhaag,
E.~von Toerne and H.~Voss,
\emph{TMVA: Toolkit for Multivariate Data Analysis},
PoS ACAT \textbf{040} (2007),
arXiv:physics/0703039.

\bibitem{Friedman2001}
J.~H.~Friedman,
\emph{Greedy Function Approximation: A Gradient Boosting Machine},
Ann. Statist. \textbf{29} (2001) 1189--1232,
doi:10.1214/aos/1013203451.


\end{thebibliography}
\end{document}